\documentclass{ifacconf}

\counterwithin*{section}{part}

\usepackage{graphicx}      
\usepackage{natbib}        
\usepackage{enumitem}
\usepackage{amsmath,amssymb,amsfonts}
\usepackage{subcaption}
\usepackage{caption}
\usepackage[capitalise]{cleveref}

\begin{document}
\begin{frontmatter}

\title{A game-theoretic, market-based approach to extract flexibility from distributed energy resources} 

\thanks[footnoteinfo]{This work was supported by the US Department of Energy under Award DOE-OE0000920 and the MIT Energy Initiative. }

\author[First]{Vineet Jagadeesan Nair} 
\author[First]{Anuradha Annaswamy} 

\address[First]{Department of Mechanical Engineering, Massachusetts Institute of Technology, Cambridge, MA, USA (e-mail: \{jvineet9,aanna\}@mit.edu)}

\begin{abstract}                
We propose a market designed using game theory to optimally utilize the flexibility of distributed energy resources (DERs) like solar, batteries, electric vehicles, and flexible loads. Market agents perform multiperiod optimization to determine their feasible flexibility limits for power injections while satisfying all constraints of their DERs. This is followed by a Stackelberg game between the market operator and agents. The market operator as the leader aims to regulate the aggregate power injection around a desired value by leveraging the flexibility of their agents, and computes optimal prices for both electricity and flexibility services. The agents follow by optimally bidding their desired flexible power injections in response to these prices. We show the existence and uniqueness of a Nash equilibrium among all the agents and a Stackelberg equilibrium between all agents and the operator. In addition to deriving analytical closed-form solutions, we provide simulation results for a small example system to illustrate our approach.
\end{abstract}


\end{frontmatter}

\section{Introduction}

In order to meet climate change mitigation goals, we need to decarbonize the power grid rapidly and clean electricity is also essential to electrify and decarbonize other sectors like heating, transportation, and heavy industry. This will entail a transition away from fossil fuels towards clean distributed energy resources such as battery storage (BS), electric vehicles (EVs), and rooftop solar photovoltaic (PV) systems. Smart inverters will enable greater controllability of inverter-based resources (IBRs) like BS and PV. In addition, we expect greater demand response capabilities through smart meters and loads like heating, ventilation, and air conditioning (HVAC) units that can provide more load flexibility. Intelligent coordination and energy management of these DERs can provide valuable grid services and help improve network operation and efficiency, in terms of metrics like operational costs, losses, and power quality through voltage and frequency regulation. However, such resources are often owned by many different independent and autonomous agents, and thus cannot be dispatched or controlled directly by operators. This motivates the use of electricity markets and price signals to control and coordinate DERs indirectly. 

\begin{figure}[htb]
  \centering
  \includegraphics[width=\columnwidth]{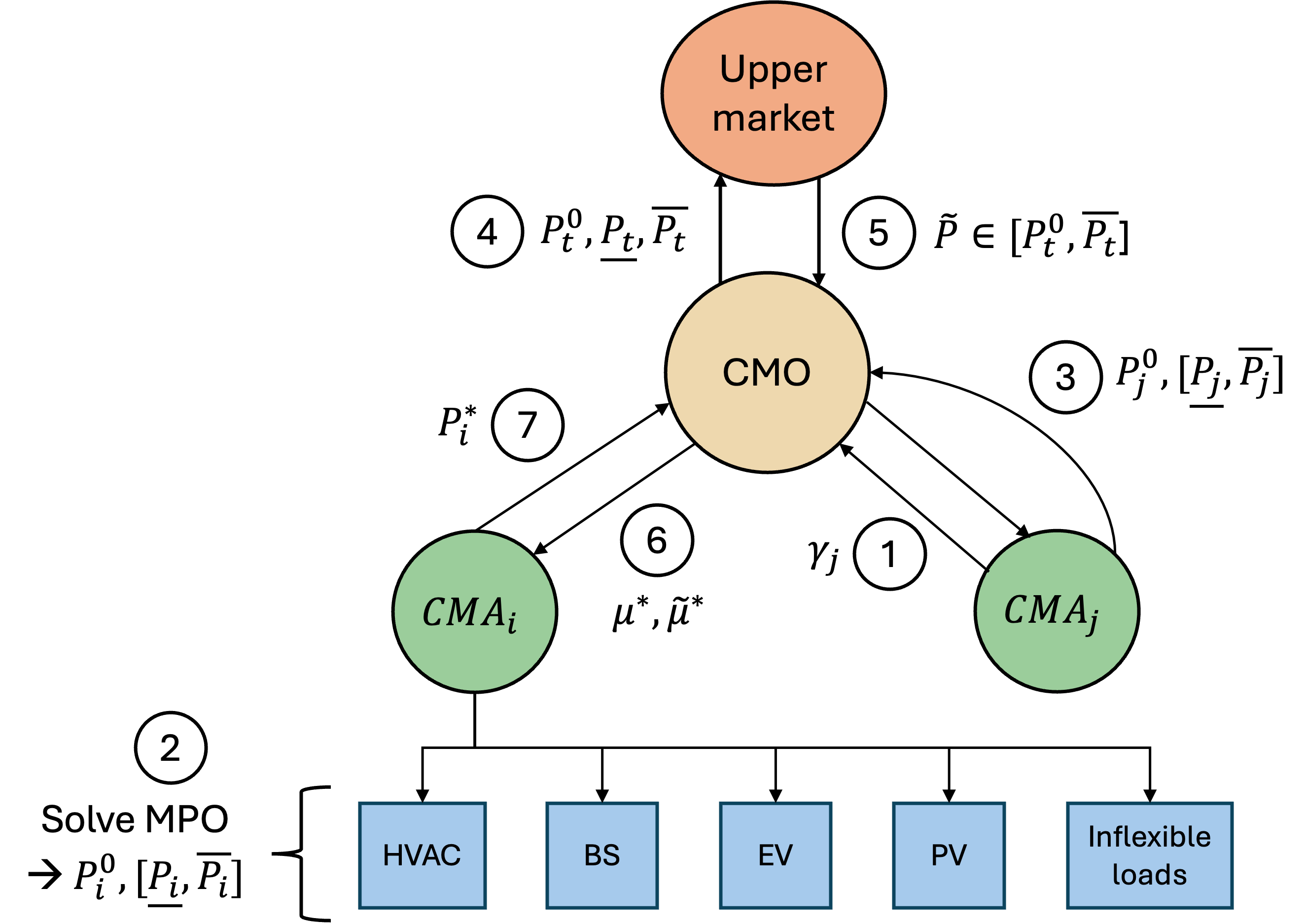}
  \caption{Overall schematic of the CM (see \cref{sec:overall_flow}). All variables will be described in the following sections.}
  \label{fig:schematic}
\end{figure}

We propose a consumer-level market (CM) structure consisting of consumer market operators (CMOs) and consumer market agents (CMAs), as seen in \cref{fig:schematic}. CMAs represent individual homes or buildings while CMOs oversee and aggregate all the CMAs under their purview. Each CMA operates several types of DERs, which offer flexibility in terms of power injections. For instance, PV can be curtailed if needed during periods of excess solar output in the middle of the day. BS and EV can be charged (or discharged) to absorb (or inject extra power). HVAC units can also provide load flexibility by varying temperature setpoints through smart thermostats. In addition to these flexible resources, we also assume that each CMA has another portion of the load that is fixed or inflexible. 

\subsection{Prior work and contributions}
Several works have proposed new kinds of local electricity markets for the future grid rich in DERs (\cite{lembook}), including retail markets (\cite{Haider2021ReinventingMarkets}) and peer-to-peer markets (\cite{sousa2019peer}). We build upon this work to analyze how game theory and mechanism design can be used to inform the development of such market structures, especially those that are closest to the end-users (electricity consumers and prosumers).

There is also rich literature studying applications of game theory (\cite{saad2012game,fadlullah2011survey}), and mechanism design (\cite{Eid2016ManagingDesign_new}) in this domain. Common modeling approaches include Stackelberg games (\cite{Maharjan2013DependableApproach} and coordinated (or coalitional) games (\cite{Turdybek2024ALoads_new,Saeian2022CoordinatedEnvironment_new}). Some works have also proposed distributed algorithms to solve such games (\cite{Li2011OptimalNetworks,Anoh2020EnergyApproach}) or utilized Vickrey–Clarke–
Groves mechanisms (\cite{Nekouei2015Game-theoreticMarkets_new}). Our work seeks to extend this by (i) more accurately modeling the physical dynamics and constraints of DERs, (ii) proposing a new approach to aggregate and maximize flexibility, and (iii) using different types of tariffs to charge or compensate agents.

\section{DER modeling \label{sec:der_models}}
Our starting point is a multiperiod optimization (MPO) problem solved by each CMA in order to determine its desired power setpoint and maximum feasible flexibility range. This MPO accounts for all the device-level constraints of individual DERs, including time-coupled state constraints for the BS, EV, and HVAC. The simulation timestep is $\Delta t$, the total simulation time is $\mathcal{T} = [0,T]$ and the planning horizon for the MPO is given by $\mathcal{H} = [t_H, t_H + (H-1)\Delta t] \subset \mathcal{T}$. We assume that the market clearing timestep is also equal to $\Delta t$. Here, $H$ is the length (number of timesteps) of the planning horizon. Note that throughout this paper, $P$ denotes net active power injections (generation minus load). Thus, $P > 0$ implies net generation while loads correspond to $P < 0$. We do not consider reactive power in this work. 

\subsection{BS model}
The state of charge (SOC) dynamics of the battery are:
\begin{gather}
    SOC^{BS}_i(t+1) = (1-\delta_{BS}^i)SOC^{BS}_i(t) - \frac{P^{BS}_i(t)\Delta t \eta^{BS}_i}{\overline{E}^{BS}_i} \label{eq:bs_soc_dynamics}\\
    \underline{P}_i^{BS} \leq P^{BS}_i(t) \leq \overline{P}^{BS}_i \label{eq:bs_p_lim}\\
    \underline{SOC}_i^{BS} \leq SOC^{BS}_i(t) \leq \overline{SOC}^{BS}_i \label{eq:bs_soc_lim}\\
    SOC_i^{BS}(0) = SOC_i^{BS}(T) \label{eq:bs_soc_terminal}
\end{gather}
where $\delta^{BS}_i, \eta^{BS}_i$ and $\overline{E}^{BS}_i$ are the BS self-discharge rate, roundtrip efficiency, and maximum capacity, respectively. We also have a terminal constraint to ensure that the state of charge at the start and end of the simulation must be equal. During BS operation, the CMA also aims to minimize the cycling cost to avoid excessive charge and discharge cycles, which can degrade the battery's lifetime.
\begin{gather}
    f^{BS}_i(P^{BS}_i) = \alpha_{cyc} \sum_{t=t_H}^{t_H + (H-1)\Delta t} (P^{BS}_i(t+1) - P_i^{BS}(t))^2
\end{gather}
where we sum over all the timesteps in the planning horizon starting at time $t_H$.

\subsection{EV model}
The EV also has SOC constraints on its battery similar to the BS. 
In addition, we place restrictions on EV availability. We assume the EV is not present at the building or home during the period $[t_1,t_2]$, e.g. between 9am and 5pm when the owner is at work.
\begin{gather}
    P_i^{EV}(t) = 0 \; \forall \; t \in [t_1, t_2] 
\end{gather}
We can impose a similar cycling cost on the EV to extend its lifetime. We also add a tracking objective that the EV owner would like to achieve a certain desired SOC ($SOC_i^*$) by a specific time ($t^*$), say the owner needs the EV to be 90\% charged by 9am before work. Thus, the EV objective function is
\begin{gather}
    f^{EV}_i = \alpha_{cyc} \sum_{t=t_H}^{t_H + (H-1)\Delta t} (P^{EV}_i(t+1) - P_i^{EV}(t))^2 \nonumber \\ 
    + \; \xi_{ev} (SOC^{EV}_i(t^*) - SOC^{EV^*}_i)^2
\end{gather}

\subsection{HVAC model}
The HVAC dynamics describe how the power drawn affects the indoor air temperature in the home or building. In this work, we consider the HVAC unit to be a heat pump (HP), which can serve as either a heating or cooling device depending on the ambient temperature. If $T^{out}_i(t) > T^{in}_i(t)$, the temperature dynamics of the HP in cooling mode (i.e. when it acts as an air conditioner) are (\cite{Zhao2017ALoads}):
\begin{gather}
    T^{in}_i(t+1) = \theta_i T_i^{in}(t) + (1-\theta_i)\left(T^{out}_i(t) + \rho_i P^{HP}_i(t)\right) \nonumber
\end{gather}
where $\theta_i = e^{\frac{-\Delta t}{R^{th}_iC^{th}_i}} \approx 1 - \frac{\Delta t}{R^{th}_i,C^{th}_i}, \rho_i = R^{th}_i \eta_i$ and $R^{th}_i, C^{th}_i, \eta_i$ are the equivalent thermal resistance, thermal capacitance, and coefficient of performance of the system, respectively. The temperature dynamics in heating mode, when $T^{out}_i(t) < T^{in}_i(t)$, are:
\begin{gather}
    T^{in}_i(t+1) = \theta_i T_i^{in}(t) + (1-\theta_i)\left(T^{out}_i(t) - \rho_i P^{HP}_i(t)\right) \nonumber
\end{gather}
The HP operation is also subject to operational limits on power draw and indoor temperature:
\begin{gather}
    \underline{P}^{HP}_i \leq P^{HP}_i(t) \leq 0 \label{eq:hvac_p_lim}\\
    \underline{T}_i^{in} \leq T^{in}_i(t) \leq \overline{T}^{in}_i \label{eq:hvac_temp_lim}\\
\end{gather}
Note that here the lower limit is determined by the maximum power consumption rating of the HVAC unit, i.e. $\underline{P}^{HP}_i = -P^{HP}_{rated,i}$ since the HP always acts as a load. The CMA would also like to track a desired temperature setpoint to maximize the thermal comfort of occupants.
\begin{gather}
    f^{HP}_i = \xi_{ac} \sum_{t=t_H}^{t_H + (H-1)\Delta t} (T^{in}_i(t) - T^{in^*}_i)^2
\end{gather}
In addition to HVAC units, using a similar approach, we may also consider other types of thermostatically controlled loads (TCLs) such as water heaters (WH).

\subsection{PV model}
The maximum PV generation output is determined by the forecasted, time-varying solar irradiance profile $\alpha^{PV}(t)$ along with its maximum rated capacity $\overline{P}^{PV}_i$, which can be curtailed if needed.
\begin{gather}
    0 \leq P^{PV}_i(t) \leq \alpha^{PV}(t) \overline{P}^{PV}_i
\end{gather}
The objective here is to minimize the amount of clean power that is curtailed.
\begin{gather}
    f^{PV}_i = \xi_{pv} \sum_{t=t_H}^{t_H + (H-1)\Delta t} (\alpha^{PV}(t) \overline{P}^{PV}_i - P^{PV}_i(t))^2
\end{gather}
In addition, the CMA would also like to utilize as much of the PV output as possible (when it's available) to charge the BS and EV, by minimizing the following objective:
\begin{gather}
    f^{util}_i = (P_i^{PV} + P_i^{BS} + P_i^{EV})^2
\end{gather}

\section{Game theoretic formulation}

We propose modeling the CM as a \textit{Stackelberg} game, where the CMO (as the leader) acts first by announcing prices and the CMAs (as followers) then respond to these prices by bidding the desired flexibility in power injections they would each like to provide. This is a game of \textit{incomplete information} (or a \textit{Bayesian} game) since CMAs do not know the private types of other players and cannot observe their actions or bids. This can also be viewed as a \textit{hierarchical} game since the CMAs are all coupled indirectly through the CMO. In addition to their flexibility bids, each player's reward (or utility) depends on the prices which in turn are influenced by others' bids as well. The CMAs compete with each other in a \textit{non-cooperative} fashion to provide flexibility to the CMO and the game is repeated at every market clearing time instance.

\subsection{CMA DER-coordination problem}

Each CMA solves an MPO to determine the optimal desired power setpoints for each DER ($P^{d^*}_i(t)$) as well as the symmetric upward or downward flexibility ($\delta_i^{d^*}(t)$) they can provide around it. In this multiobjective optimization, the CMA aims to (i) maximize the total flexibility it can provide, (ii) maximize the PV utilization for charging, (iii) minimize the costs associated with providing flexibility (as specified in \cref{sec:der_models}), and (iv) maximize the total net power injection into the grid i.e. maximize net generation (import) or minimize net load (import). This is subject to all the DER-level constraints and dynamics described in the preceding sections in addition to operational upper and lower limits with flexibility. The MPO formulation is:
\begin{gather}
    \min_{P_i^d(t), \delta_i^d(t)} \sum_{t \in \mathcal{H}} \sum_{d \in \mathcal{D}} -\delta_i^d(t) + f_i^d(P_i^d) + f_i^{util}(P_i^d) - P^{total}_i(t) \label{eq:cma_mpo} \\
    \text{s.t. } \underline{P^d_i(t)} \leq P_i^d(t) - \delta_i^d(t), \; P_i^d(t) + \delta_i^d(t) \leq \overline{P^d_i(t)} \nonumber \\
    \text{All device-specific state constraints for each DER } d \in \mathcal{D}_i \nonumber\\
    P^{total}_i(t) = \sum_{i \in \mathcal{D}_i} P^d_i(t) - P^{fixed}_i(t), \; \epsilon_1 |P^d_i| \leq \delta^d_i \leq \epsilon_2 |P^d_i| \nonumber
\end{gather}
where $\mathcal{D}_i \subseteq \{BS,EV,HVAC,PV\}$ is the set of all DERs owned and managed by CMA $i$. Note that we also place upper and lower limits on the flexibility of each DER (with $\epsilon_1 < \epsilon_2$). Since the absolute values are non-convex, we used an exact big-M reformulation to represent $|P^d_i| = P^{d,abs}_i$ by introducing additional binary variables. Note that we only need to do this $d = BS, EV$ since for the other devices we know $|P^{HP}_i| = -P^{HP}_i$ and $|P^{PV}_i| = P^{PV}_i$.
\begin{gather}
P^{d}_i=P^{d,+}_i-P^{d,-}_i, \; |P^{d}_i|=P^{d,+}_i+P^{d,-}_i, \; z^d_i \in \{0,1\} \nonumber \\
0 \leq P^{d,+}_i \leq z^d_i \overline{P^d_i}, \; 0 \leq P^{d,-}_i \leq (1-z^d_i)|\underline{P^d_i}|
\end{gather}
Thus, the overall MPO is a mixed integer quadratic program (MIQP). At each market clearing $t$, after solving the MPO in \cref{eq:cma_mpo}, the CMA aggregates the solutions and flexibilities across all their DERs to determine their total baseline power injections $P^0_i$ and feasible flexibility range $[\underline{P}_i,\overline{P}_i]$ that they could provide to the CMO. Note that this range is their maximum possible flexibility based on device-level costs and physical constraints of DERs. It does not account for the CMA's utility, welfare, or strategic behavior. This will be considered in \cref{sec:cma_game}.
\begin{gather}
    P^0_i = \sum_{d \in \mathcal{D}_i} P^{d^*}_i, \; \underline{P}_i = P^0_i - \sum_{d \in \mathcal{D}_i} \delta_i^{d^*}, \; \overline{P}_i = P^0_i + \sum_{d \in \mathcal{D}_i} \delta_i^{d^*} \nonumber
\end{gather}

\subsection{CMA welfare maximization problem for game \label{sec:cma_game}}
Having determined $P^0_i, [\underline{P}_i,\overline{P}_i]$, the CMA solves the following constrained optimization to maximize its social welfare $U^{cma}_i$ and thus determine its optimal bid into the CM. We consider the CMA as a for-profit, strategic aggregator of DERs.
\begin{gather}
    \max_{P_i} U^{cma}_i(P_i, \widetilde{\mu},\mu) = \widetilde{\mu} (P_i - P_i^0) + \mu P_i - \gamma_i (P_i-P_i^0)^2  \nonumber \\
    \text{s.t. } \underline{P}_i \leq P_i \leq \overline{P}_i, P_i \geq P_i^0 \label{eq:cma_game_opt}
\end{gather}
where $\gamma_i > 0$ is the CMA's disutility preference parameter and denotes their type. We assume in this work that the CMO always requests only \textit{upward} flexibility in net power injections from the CMAs, i.e., increasing net injections either by reducing demand (via load shifting or curtailment) or increasing generation. This is in line with demand response programs today. This ensures that the first term in the objective function is always non-negative, since $(P_i - P_i^0) \geq 0$. This first term is the compensation given by the CMO to CMAs for any flexibility they provide to the market, at the rate of $\widetilde{\mu}(t)$. The second term represents the net cost to the CMA, which constitutes payments to the CMO (in case the CMA is a net load $P_i < 0$) or revenue received from the CMA (in case the CMA is a net generator $P_i > 0$). This depends on the electricity price $\mu(t)$. The last term is the disutility caused to the CMA by flexibility, due to deviations from its ideal (nominal) injection $P_i^0$. Thus the CMA aims to maximize its flexibility compensation and minimize disutility and net costs while satisfying the injection limits from stage I. Note that $U^{cma}_i$ is concave for the maximization problem, thus the corresponding minimization problem is a convex quadratic program (QP).

In this game, each CMA's action or strategy is given by their flexible power injection bid $P_i^*(t)$. The baseline electricity price $\mu(t)$ and flexibility price $\widetilde{\mu}(t)$ are determined by the CMO's strategy. Note that the prices vary over time but are common for all CMAs. By applying the Karush–Kuhn–Tucker (KKT) conditions, we can analytically derive the optimal solutions for \cref{eq:cma_game_opt}:
\begin{gather}
    P_i^*(\widetilde{\mu},\mu) = \begin{cases}
  \overline{P}_i & \text{if } \frac{\gamma_i(\overline{P}_i - P_i^0)}{\widetilde{\mu} + \mu} < \frac{1}{2} \\
  P_i^0 + \frac{\widetilde{\mu}+\mu}{2\gamma_i} & \text{otherwise}
\end{cases} \label{eq:cma_game_soln}
\end{gather}

\subsection{CMO optimization problem}

The goal of the CMO is to track the setpoint $\widetilde{P}(t)$ for the total power injection, as determined by the market above it. It does so by leveraging the flexibility bids of their CMAs. The CMO thus aims to extract as much flexibility as possible from the CMAs while accounting for their preferences and managing costs. The CMO can then offer this aggregate flexibility to the higher-level market. This upper market could be a local electricity market (LEM) for the distribution grid, as proposed in our previous work (\cite{Nair2022AEdge}). Alternatively, the CMO could also participate directly in the wholesale energy market (WEM) by bidding its flexibility at the main transmission grid level. It has to respect the power bounds of each CMA as specified by the bids in addition to a budget balance. This requires that the CMO exactly breaks even at each market clearing time step. We assume here that the CMO is a not-for-profit entity that serves just as a coordinator, but we will generalize to the case where the CMO is profit-making as part of future work. 

At each market clearing $t$, given the flexible power injection bids $P^*_i(t)$ and the other inputs $P^0_i(t), [\underline{P}_i(t),\overline{P}_i(t)]$ from all CMAs $i$, the CMO solves the following feasibility optimization problem to match the desired $\widetilde{P}$ and set both the prices to satisfy the budget balance constraint while respecting each CMA's flexibility bids. It's crucial to carefully set these prices so as to encourage the desired power injections from the CMAs, while accurately valuing and compensating any flexibility they provide. The budget constraint also ensures that the CMO meets its flexibility commitment $\widetilde{P}$ in the most cost-effective manner. We assume a lossless power balance between the CMO and CMAs, as well as between the CM and the LEM above.
\begin{gather}
    \min_{\widetilde{\mu}(t),\mu(t)} -U^{cmo}(P_i(t),\mu(t),\widetilde{\mu}(t)) = \left(\sum_{i\in \mathcal{C}} P_i(t) - \widetilde{P}(t)\right)^2 \nonumber \\
    \text{s.t. } P_i^0(t) \leq P_i(t) \leq \overline{P}_i(t) \; \forall \; i \label{eq:cmo_opt}\\
    \widetilde{\mu}(t)\sum_{i \in \mathcal{C}} (P_i(t)-  P_i^0(t)) + \mu(t) \sum_i P_i(t) = \pi(t) \sum_i P_i(t) \nonumber
\end{gather}
Where $\mathcal{C}$ is the set of all CMAs under the purview of the CMO, and $\pi(t)$ is the time-varying electricity price set by the LEM above. This would be the locational marginal price (LMP) if the CMO participates directly in the WEM. Alternatively, this could be the d-LMP or retail rate if the CMO is part of another market in the distribution grid. The budget balance constraint implies that the net cost to the CMO for running the CM (i.e. total payment to CMAs) must equal the net revenue it receives from the LEM, or conversely, the net total revenue it receives from all the CMAs must equal the total payment it makes to the LEM. Summing up terms n the budget balance in \cref{eq:cmo_opt} over all CMAs $i$ gives:
\begin{gather}
    \widetilde{\mu}(P_t - P_t^0) + \mu P_t = \pi P_t \label{eq:cmo_game_budget_sum}
\end{gather}
where $P_t = \sum_i P_i$ and $P^0_t = \sum_i P_i^0$. Further, we assume that at the start of each market operation timestep (i.e. before price scheduling and bidding), the CMAs directly report their solutions from the MPO (\cref{eq:cma_mpo} to the CMO, i.e. $P_i^0$ and feasible $[\underline{P}_i(t),\overline{P}_i(t)]$. The CMO then aggregates these and reports the total baseline injection $P^0_t$ as well as the upper and lower bound of total available flexibility $\underline{P}_t = \sum_i \underline{P}_i, \overline{P}_t = \sum_i \overline{P}_i$ to the upper LEM. This ensures that the LEM gives a command for the desired power setpoint that is feasible for the CM to provide, i.e., $\underline{P}_t \leq \widetilde{P} \leq \overline{P}_t$. We further assume that the LEM may only request upward flexibility implying that $\widetilde{P} \geq P^0_t$. $\widetilde{P}$ and $P^0_t$ will also have the same sign since a net load (or generator) will remain a net load (or generator) even after flexibility provision.

\subsection{Computing optimal prices}
We now show how the CMO can set optimal prices for electricity and flexibility. Firstly, we note that CMO always sets prices such that $\mu + \widetilde{\mu} < 2\gamma_i (\overline{P}_i - P_i^0) \; \forall \; i$. According to \cref{eq:cma_game_soln}, it follows that $P_i^* \neq \overline{P}_i \; \forall \; i$. The CMO sets this condition to ensure that no single CMA $i$ is disproportionally affected and we avoid relying excessively on any of the CMAs for providing flexibility. Thus, summing up the optimal flexible power injections across all CMAs in \cref{eq:cma_game_soln}, we get the following
\begin{gather}
    P_t^* = \sum_{i \in \mathcal{C}} P_i^* = P^0_t + \gamma_t \frac{\mu + \widetilde{\mu}}{2} = \widetilde{P} \label{eq:cma_game_soln_sum}
\end{gather}
where $\gamma_t = \sum_{i \in \mathcal{C}} \frac{1}{\gamma_i}$ and we also used the fact that at the optimum of \cref{eq:cmo_opt}, we have $P_t^* = \widetilde{P}$ since we know the setpoint $P^0_t \leq \widetilde{P} \leq \overline{P}_t$ is feasible to achieve for the CMO. The CMO now needs to compute the optimal $\mu^*, \widetilde{\mu}^*$ that cause all the CMAs to bid $P_i^*$ such that $\sum_i P_i^* = \widetilde{P}$ exactly. Combining \cref{eq:cmo_game_budget_sum} and \cref{eq:cma_game_soln_sum}, we can derive the baseline electricity price:
\begin{gather}
    \mu^* = \frac{\pi \widetilde{P}}{P^0_t} - \frac{2(\widetilde{P}-P^0_t)^2}{\gamma_t P^0_t} \label{eq:mu_soln} 
\end{gather}
Based on \cref{eq:cmo_game_budget_sum}, we can then also obtain the flexibility price based on the electricity tariff as follows:
\begin{align}
    \widetilde{\mu}^* & = \frac{P_t^* (\pi - \mu^*)}{P_t^* - P_t^0} = \frac{\widetilde{P} (\pi - \mu^*)}{\widetilde{P} - P_t^0} = \frac{\widetilde{P}\left(2(\widetilde{P}-P^0_t) - \pi \gamma_t\right)}{\gamma_t P^0_t}\label{eq:mu_tilde_soln} 
\end{align}

\subsection{Positivity of prices}
We also need to verify certain conditions to ensure that the prices $\mu^*(t), \widetilde{\mu}^*(t) > 0 \; \forall \; t$. From \cref{eq:mu_tilde_soln}, we see that:
\begin{gather}
    \widetilde{\mu}^* > 0 \implies \frac{\widetilde{P}\left(2(\widetilde{P}-P^0_t) - \pi \gamma_t\right)}{\gamma_t P^0_t} > 0 \label{eq:mu_tilde_+ve_cond} \\
    \implies 2(\widetilde{P}-P^0_t) - \pi \gamma_t > 0 \; \because \frac{\widetilde{P}}{P^0_t} >0  \implies \widetilde{P} > P^0_t + \frac{\pi \gamma_t}{2} \nonumber
\end{gather}
Given that $\widetilde{P} > P_t^0$, we note that this condition generally holds true for most realistic values of $\gamma_t$, and $\pi$ since the magnitude of the term $\frac{\pi \gamma_t}{2}$ is small relative to $P_t^0$. 

From \cref{eq:mu_soln}, we can also derive conditions for $\mu^* >0$:
\begin{gather*}
    \frac{\pi \widetilde{P}}{P^0_t} - \frac{2(\widetilde{P}-P^0_t)^2}{\gamma_t P^0_t} > 0 \implies \frac{\pi \widetilde{P}\gamma_t - 2(\widetilde{P}-P^0_t)^2}{\gamma_t P^0_t} >0    
\end{gather*}
We notice that $\mu^* > 0$ for any negative values of $\widetilde{P}, P^0_t$, i.e. if the CMO as a whole is a net load. However, if the CMO is a net generator, i.e., $\widetilde{P}, P^0_t > 0$, we require the following conditions to hold true:
\begin{gather}
    \mu^* >0 > \implies (\widetilde{P} - P^0_t)^2 < \frac{\pi \widetilde{P} \gamma_t}{2}
    \implies a_1 < \widetilde{P} < a_2 \nonumber\\
    a_1 = P_t^0 + \frac{\pi \gamma_t}{4} - \frac{1}{2}\sqrt{\frac{\pi \gamma_t}{2}\left(4P_t^0 + \frac{\pi \gamma_t}{2}\right)} \nonumber \\
    a_2 = P_t^0 + \frac{\pi \gamma_t}{4} + \frac{1}{2}\sqrt{\frac{\pi \gamma_t}{2}\left(4P_t^0 + \frac{\pi \gamma_t}{2}\right)} \label{eq:mu_+ve_cond} 
\end{gather}
Combining \cref{eq:mu_tilde_+ve_cond} and \cref{eq:mu_+ve_cond}, we get the following on $\widetilde{P}$ for positivity of $\mu^*, \widetilde{\mu}^*$:
\begin{gather}
\begin{cases}
  \widetilde{P} > P^0_t + \frac{\pi \gamma_t}{2} & \text{if } \widetilde{P}, P^0_t < 0 \\
  \max\left(a_1, P_t^0 + \frac{\pi \gamma_t}{2}\right) \leq \widetilde{P} \leq a_2 & \text{if } \widetilde{P}, P^0_t > 0 \label{eq:mu_+ve_cond_combined}
\end{cases}
\end{gather}
During our numerical simulations, the CMO as a whole was generally a net load (i.e. $\widetilde{P}, P_t^0 < 0$) and we found that the conditions above in \cref{eq:mu_+ve_cond_combined} consistently held true for all time periods, ensuring that the prices remained positive throughout the day. 

\subsection{Overall process for market operation\label{sec:overall_flow}}
We summarize the steps for our CM operation as follows (also see \cref{fig:schematic}):
\begin{enumerate}
    \item All CMAs report their type $\gamma_i$ to the CMO only at the start of the game and these remain the same for all subsequent market clearings. In this work, we assume that the CMAs report their type truthfully. 
    
    \item CMAs solve MPO in \cref{eq:cma_mpo} to determine ideal baseline injection $P_i^0$ and maximum feasible flexibility intervals $[\underline{P}_i,\overline{P}_i]$. 
    
    \item CMAs communicate their baseline power injection $P_i^0$ and flexibility range $[\underline{P}_i, \overline{P}_i]$ to the CMO. We argue that the CMAs are incentivized to be truthful about $P_i^0$ to maximize their utility (see \cref{eq:cma_game_opt}) since this is their ideal power injection if no flexibility is needed.

    \item The CMO reports aggregated $P^0_t, \underline{P}_t, \overline{P}_t$ to the LEM.

    \item The LEM requests a feasible flexible setpoint $\widetilde{P} \in [P^0_t,\overline{P}_t]$ from the CMO.

    \item The CMO announces optimal prices $\mu^*, \widetilde{\mu}^*$ to all its CMAs, to track the regulation signal $\widetilde{P}$.
    
    \item CMAs respond to the prices with their optimal bids for desired flexible power injections $P_i^*$, following which the market is cleared and settled.
\end{enumerate} 

We note here that we still need to verify the incentive compatibility (IC) of our market design to ensure that the CMAs report their true types and bid according to their actual preferences and flexibility capabilities. We will study this more rigorously as part of future work. Moreover, we know from the revelation principle that it is possible, if needed, to come up with a truthful, IC mechanism to implement our optimal bidding and pricing functions (derived above) and achieve the same Bayesian equilibrium outcome and welfare.

\subsection{Equilibrium}
The optimal prices $\mu^*, \widetilde{\mu}^*$ set by the CMO will induce optimal bids $P_i^*$ from all CMAs that lead to an equilibrium in pure strategies.
\begin{thm}
The set of bids and prices $\{P_i^* \; \forall \; i, \mu^*, \widetilde{\mu}^*\} $ corresponds to a unique Nash equilibrium (NE) amongst the CMAs and thus also a unique Stackelberg equilibrium (SE) between all the CMAs and the CMO.
\end{thm}
\begin{pf}
    Note that the welfare of each CMA $i$ also depends others' actions and are coupled through the prices:
    \begin{gather}
        U^{cma}_i(P_i, \widetilde{\mu}(P_i, P_{-i}), \mu(P_i, P_{-i})) \equiv U^{cma}_i(P_i, P_{-i}, \widetilde{\mu}, \mu) \nonumber
    \end{gather}
    where $P_{-i}$ denotes the flexible power injection bids of all the CMAs other than CMA $i$. Given the electricity and flexibility prices $\mu^*, \widetilde{\mu}^*$ set by the CMO, the CMAs bid their welfare-maximizing flexible power injections $P_i^*$ according to \cref{eq:cma_game_soln}. Thus, this strategy profile $\{P_i^*\}$ corresponds to a unique NE among the CMAs since $\forall P_i \in [\underline{P}_i,\overline{P}_i]$, we have:
    \begin{gather}
        U^{cma}_i(P_i^*, P_{-i}^*, \widetilde{\mu}^*, \mu^*)\geq U^{cma}_i(P_i, P_{-i}^*,\widetilde{\mu}^*, \mu^*) \nonumber
    \end{gather}
    Similarly, given that all the CMAs submit their best response flexible injection bids $P_i^*$, the optimal prices $\mu^*, \widetilde{\mu}^*$ from \cref{eq:mu_soln,eq:mu_tilde_soln} result in a unique SE between the CMO and the CMAs. This is because $\mu^*, \widetilde{\mu}^*$ is the unique pair of prices that balances the budget constraint in \cref{eq:cmo_opt} and exactly matches the total power injection $\sum_i P_i$ to equal $\widetilde{P}$, thus implying that:
    \begin{gather}
        U^{cmo}(P_i^*,P_{-i}^*,\mu^*,\widetilde{\mu}^*) = 0 \geq U^{cmo}(P_i^*,P_{-i}^*,\mu,\widetilde{\mu}) \; \forall \mu, \widetilde{\mu} \nonumber
    \end{gather}
\end{pf}
\textit{Note}: Nash's theorem states the existence of mixed strategy Nash equilibrium (NE) for any finite game (with discrete strategy sets) and Glicksberg's theorem extends this to the infinite setting with continuous strategies, as in our case (\cite{glicksberg1952further}). However, in reality, computing these equilibria can be very challenging. We thus needed to make several assumptions and simplifications to be able to readily show the existence and uniqueness of the NE, and derive it analytically.

\section{Simulation results}
\begin{figure}[htb]
\centering
\begin{subfigure}[t]{0.49\columnwidth}
  \includegraphics[width=\columnwidth]{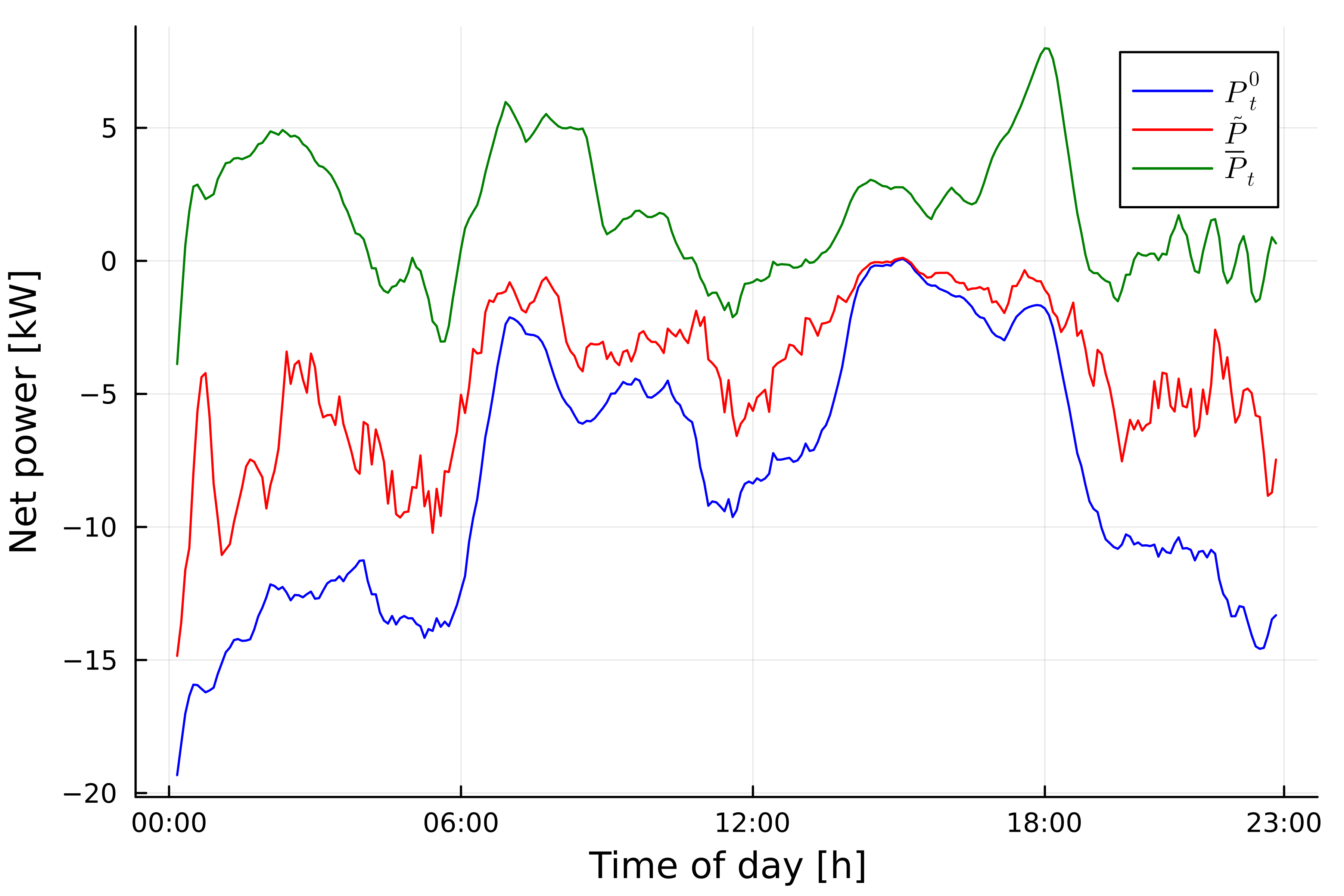}
  \caption{Total CMO power injections.}
  \label{fig:cmo_total_power}
\end{subfigure}
\begin{subfigure}[t]{0.49\columnwidth}
  \includegraphics[width=\columnwidth]{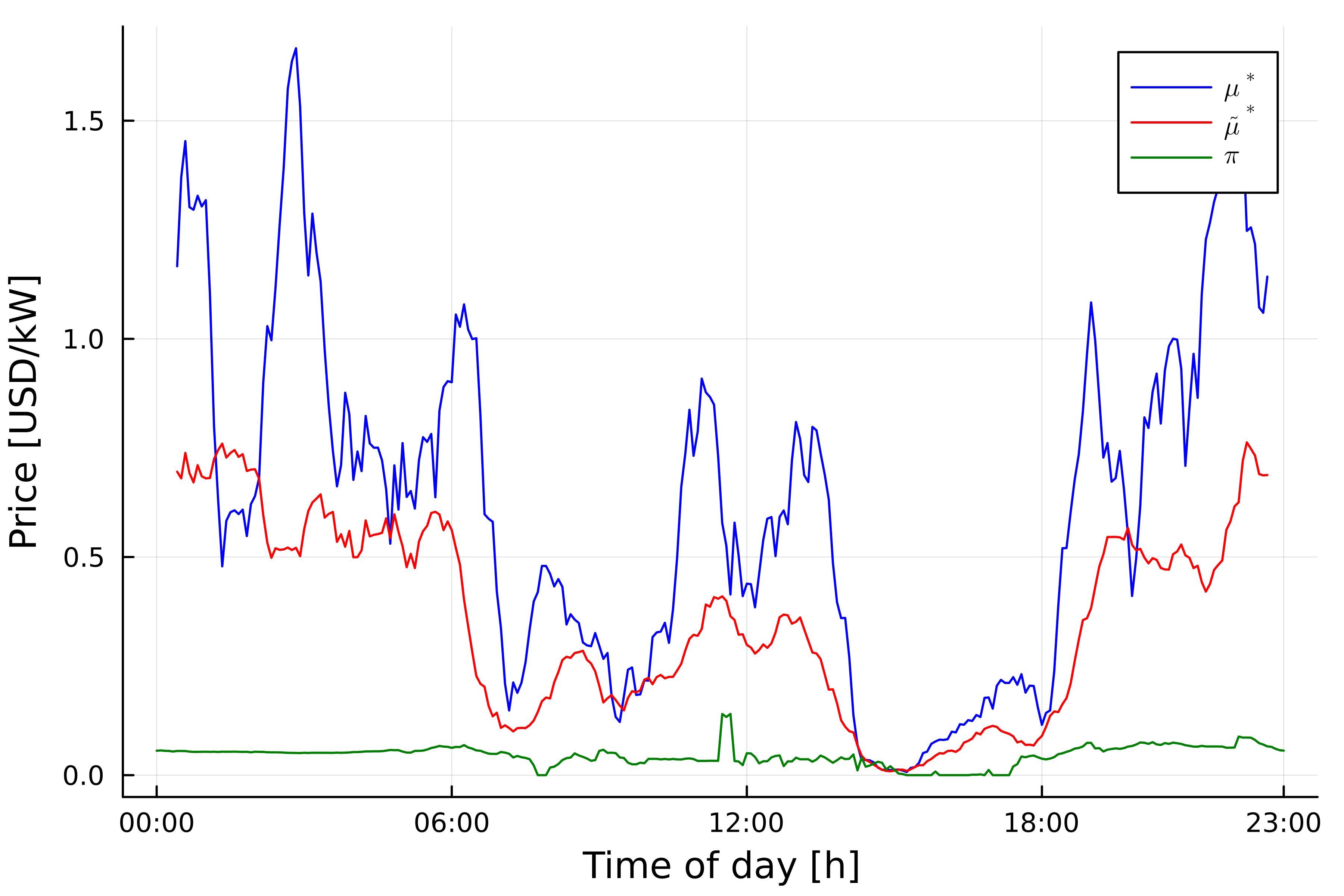}
  \caption{Prices over the day.}
  \label{fig:prices}
\end{subfigure}
\caption{CMO power injections and prices.}
\end{figure}

We now briefly describe some simulation results. We simulated a hypothetical CM with 1 CMO and 3 CMAs, each of whom has all the types of DERs (BS,EV,HVAC,PV) along with some fixed load. We used outdoor air temperature and solar PV production data for San Francisco, California, and price data for $\pi(t)$ from CAISO. \cref{fig:cmo_total_power} shows the total power injections for the CMO, aggregated over all 3 CMAs. We see that the CMO remains a net load throughout the day, with the net load lower mid-day during peak solar PV output. The LEM requests varying amounts of load flexibility or demand response throughout the day. \cref{fig:prices} shows the original LEM price $\pi$ along with the CM electricity price $\mu^*$ and flexibility price $\mu^*$. We notice that the CM prices are about an order of magnitude higher than the LEM price. Thus, for the CMO to provide the required flexibility to the LEM, it has to increase its prices and also compensate CMAs and DERs sufficiently, which raises costs. We are currently exploring approaches to mitigate these price impacts. One possibility is to use varying prices $\mu_i, \widetilde{\mu}_i$ for each CMA $i$ rather than a common rate. Although this makes the equilibrium analysis more challenging, it could lead to more equitable and fair tariffs (\cite{Cohen2022PriceConstraints_new}).

\begin{figure}[htb]
\centering
\begin{subfigure}[t]{0.49\columnwidth}
  \includegraphics[width=\columnwidth]{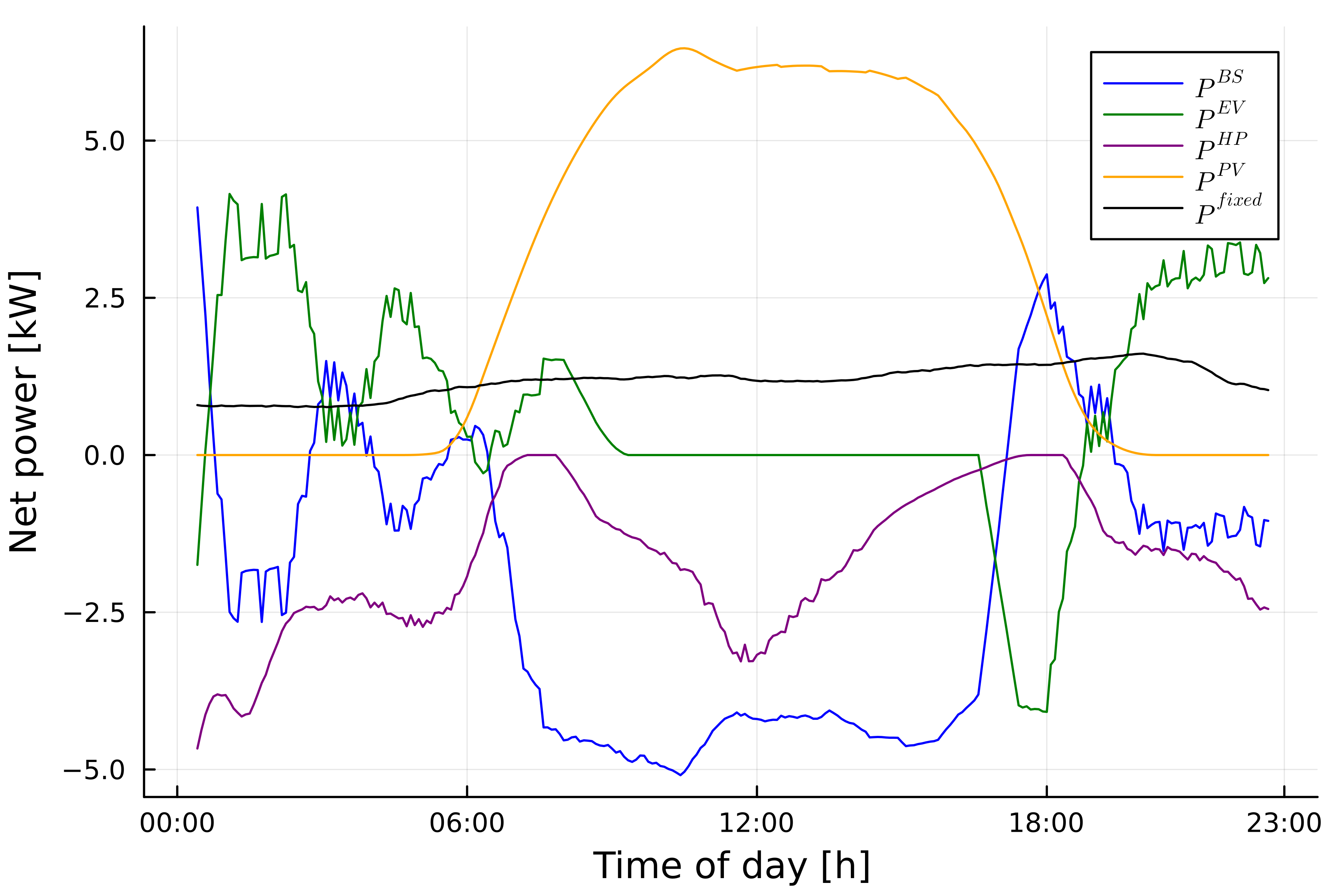}
  \caption{DER power injections.}
  \label{fig:der_injections}
\end{subfigure}
\begin{subfigure}[t]{0.49\columnwidth}
  \includegraphics[width=\columnwidth]{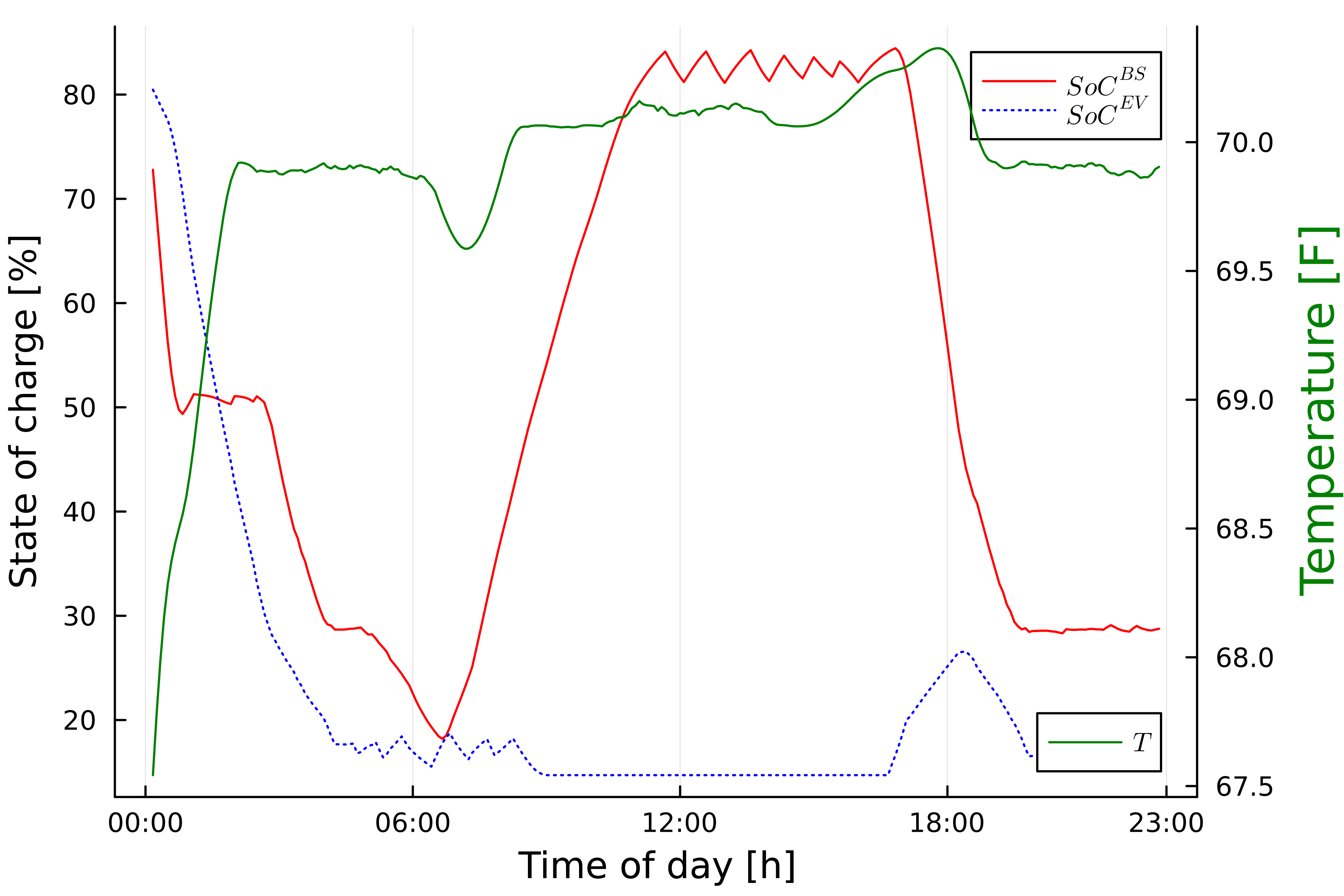}
  \caption{Temperature and state of charge profiles.}
  \label{fig:cma_states}
\end{subfigure}
\caption{DER power injections and states for CMA 1.}
\end{figure}

\cref{fig:der_injections} shows the power injections of various DERs owned by CMA 1 over the course of the day. As expected intuitively, we see that the BS charges using the solar PV output during the middle of the day, while the EV is only available to charge over a small fraction of this period. On the other hand, both the BS and EV generally discharge through other parts of the day (evening through early morning) to provide significant flexibility. These are also reflected in the SOC profile of the BS and EV over the day, shown in \cref{fig:cma_states}. We see that the power consumption of the HVAC unit or heat pump (HP) also varies quite significantly throughout the day, confirming that temperature setpoint control can indeed provide load flexibility to the CMA, while still maintaining the temperature profile close to the desired comfortable value of 70 F, as seen in \cref{fig:cma_states}. The HP generally serves as a cooling device throughout the day except from late night to early morning when it acts as a heater instead due to lower outdoor temperatures. 

\section{Conclusions and future work}
In this paper, we proposed a consumer-level market structure for a CMO to coordinate and aggregate several CMAs, each of whom operates several DERs. In our hierarchical structure, the CMAs first coordinate their DERs to determine their flexibility capabilities while accounting for all device-specific constraints. This is followed by a Stackelberg, incomplete information game between the CMO who sets prices, and CMAs who respond to the prices with flexibility bids. We derive analytical solutions for prices and bids that lead to an equilibrium among all market participants, along with simulation results for a small instantiation of the CM.

As part of future work, we will conduct simulations on larger systems. We also plan to extend our work to more realistic settings by relaxing some of our assumptions about common knowledge and information availability. To do so, we will leverage more advanced game-theoretic approaches and possibly other notions of equilibria. In more realistic and complex settings, we will also explore numerical approaches to compute approximate equilibrium solutions.

\bibliography{references,refs_manual_new}

\end{document}